\documentclass{article}

\usepackage{lineno}
\usepackage{hyperref}
\hypersetup{colorlinks,allcolors=black}
\usepackage[UKenglish,USenglish]{babel}
\usepackage{float}
\usepackage{needspace}
\usepackage{tabularx}
\usepackage{mathcomp}
\usepackage{gensymb}
\usepackage[utf8]{inputenc}
\usepackage[parfill]{parskip}
\usepackage{indentfirst}
\usepackage[numbers,sort&compress]{natbib}
\usepackage{amsmath}

\usepackage{lettrine}
\usepackage{amsthm}
\usepackage{multirow}
\usepackage{tabularx}
\usepackage{diagbox}
\usepackage[linesnumbered,ruled]{algorithm2e}
\usepackage{tikz}
\usepackage{booktabs}
\usepackage{mdframed}
\usepackage[most]{tcolorbox}
\usepackage{psfrag}
\usepackage{caption}
\usepackage{varwidth}
\usepackage{authblk}
\begin{document}

\author{ Sreyash Sarkar\thanks{corresponding author: sreyash.sarkar@esiee.fr}}
\author{Elyes Nefzaoui\thanks{corresponding author: elyes.nefzaoui@esiee.fr}}
\author{Philippe Basset}
\author{Tarik Bourouina}
\affil{ESYCOM, Univ Gustave Eiffel, CNRS, CNAM, ESIEE Paris, F-77454 Marne-la-Vallée, France}
\title{Far-field Radiative Thermal Rectification with Bulk Materials}

\maketitle

\begin{abstract}
In this paper, we explore the far-field radiative thermal rectification potential of common materials such as metals, dielectrics and doped semi-conductors using radiative and thermo-radiative properties extracted from literature. Ten different materials are considered. The rectification coefficient is then calculated for 44  pairs of materials; each pair can be used for the two terminals of a radiative thermal diode. A thermal bias of 200 K is considered. The choice of materials and thermal bias value are only bound by data availability in literature. Obtained results, highlight new candidate materials for far-field radiative thermal rectification. They also highlight materials where thermal rectification is not negligible and should be considered with care in heat transfer calculations when considering systems subject to a comparable thermal bias and where these materials are used. Among the materials studied, undoped Indium Arsenide (InAs) shows great promise to be employed for thermal rectification, with a thermal rectification ratio reaching 90.5\% in combination with other materials. Obtained results pave the way for an optimized design of thermal radiation control and management devices such as thermal diodes. 
\end{abstract}

\section{Introduction}
Thermal rectification (TR) can be defined as an asymmetry in the heat flux when the temperature difference between two interacting thermal reservoirs is reversed. Therefore, a two-terminal thermal device exhibits thermal rectification if it transports heat in one direction with more ease than in the reverse direction. Thermal rectification has been a subject of intrigue since its underlying test perception in 1936 by Starr \citep{starr_copper_1936} because of its aptitude to open up innovation in heat transport control, inspiring analogies to the tremendous advancement in the electronics industry following the invention of such nonlinear elements as the transistor and the diode. Other potential applications of thermal rectification include thermal barrier coatings \citep{miller2009thermal}, enhanced efficiency thermoelectric devices \citep{saha2006monte,benenti2016thermal}, or temperature variation driven heat engines\citep{sothmann2012rectification}. From that point forward, several investigations have been performed to understand which systems can exhibit thermal rectification\citep{roberts_review_2011} and have introduced the concepts of numerous innovative devices like thermal transistors \citep{ordonez2019radiative,ben2014near,joulain2015modulation, joulain2016quantum}, thermal logic circuits\citep{christoph2020scalable, casati2007device, murad2013thermal,nefzaoui2015tunable, kathmann2020scalable} and thermal diodes\citep{martinez2015rectification,ben2013phase,ordonez2017photonic, ott2019radiative,ordonez2018conductive, chen_photon_2014}.  
Recent  investigations of thermal rectification have covered different heat transfer(HT) modes including conduction \citep{zhang2011thermal,avanessian2015adsorption}, convection \citep{reis1987convective} and radiation \citep{rogers1961heat,clausing1966heat,lewis1968heat,yang2008carbon,li_thermal_2004,otey_thermal_2010,francoeur_electric_2011,iizuka_rectification_2012}. We describe the main current trends in the following paragraphs and more extensive reviews can be found in \citep{roberts_review_2011,ben-abdallah_contactless_2015}.

To achieve conductive thermal rectification, several mechanisms have been proposed including thermal potential barrier between material contacts\citep{rogers1961heat}, thermal strain/warping at interfaces of two materials\citep{clausing1966heat}, nanostructured geometric asymmetry\citep{shen2018high} and anharmonic lattices\citep{li_thermal_2004}. 
The asymmetry at the interface between two materials due to the difference of their thermal conductivity temperature dependence has also been shown to be the main driving mechanism for conductive thermal rectification by Marucha et. al\citep{marucha1975heat} and subsequently used in the models of  Hoff et. al\citep{hoff1985asymmetrical}, Sun et. al\citep{sun2001evaluation} and in Go et. al\citep{go2010condition}. Based on the same principle,  Hu et al\citep{hu2009thermal} and Zhang et. al\citep{zhang2011thermal}  presented thermal rectifiers based on different thermal conductivities of dissimilar graphene nano-ribbons or Y junctions. 
At the micro and nano-scales, the asymmetry at the origin of TR can be widely tuned and engineered in nano-structured materials \citep{yang2018effect} which led to many contributions using several phenomena observed in a large diversity of nano-structures such as ballistic phonons anisotropy at large thermal bias \citep{miller2009thermal}, phonons lateral confinement\citep{wang2014phonon} and the temperature dependence of lattice vibrations density of states\citep{lee2012thermal}.
Solid-state thermal rectifiers have also been proposed taking benefit of nonlinear atomic vibrations\citep{li_thermal_2004}\citep{chang_solid-state_2006}, nonlinear dispersion relations of the electron gas in metals\citep{segal2008single}, and thermal streams through Josephson junctions\citep{martinez-perez_efficient_2013} or metal-superconductor nano-intersections\citep{giazotto2013thermal}.
At even smaller scales, TR in quantum systems has recently lured researchers  such as in the work of Landi et.al\citep{landi2014flux} on TR using a quantum XXZ chain subject to an inhomogeneous field and  the research of Chi et. al \citep{landi2014flux} discussing TR in a system of a single level quantum-dot connected to ferromagnetic leads. On the other hand, Scheibner et al\citep{scheibner_quantum_2008} experimentally demonstrated a  TR behavior in quantum dots subject to high in-plane magnetic fields\citep{scheibner_quantum_2008} and most recently in the work of Senior et. al\citep{senior2020heat}, it has been shown that a superconducting quantum bit coupled to two superconducting resonators can achieve magnetic flux-tunable photonic heat rectification between 0 and 10\%. 

Radiative heat transfer is another heat transfer mode that has been actively investigated for thermal rectification application in the past decade after the seminal works of Ruokola et. al\citep{ruokola2009thermal} and Otey \citep{otey_thermal_2010}.
Two main paths have been considered. 
On one hand, near field (NF) Radiative Thermal Rectification (RTR) between materials supporting resonating surface waves spearated by wavelengeth scale or sub-wavelength gaps. Plasmonic materials supporting surface plasmon polaritons and dieletrics suppporting surface phonon polaritons have been mainly considered.
Used materials for NF RTR include gold and silicon \citep{wang_thermal_2013,wen2019ultrahigh} with temperatures between \(300 K\) and \(600 K\), doped silicon\citep{basu_near-field_2011}, SiC  \citep{otey_thermal_2010,francoeur_electric_2011,zhu2013ultrahigh}, SiC and SiO\(_{2}\) \citep{joulain_radiative_2015, iizuka_rectification_2012} , InSb and graphene-coated SiO2\citep{xu2018highly}. Many works on NF RTR  also used phase change materials and mainly Vanadium Dioxyde (\(VO_2\))\citep{ghanekar2016high, zheng2017graphene, fiorino2018thermal, gu2015thermal}, and more rarely, thermochromic materials \citep{huang_thermal_2013}.
In addition to the diversity of used materials, a large variety of geometries and sizes has been explored as in the work of Shen et. al\citep{shen2018high}, which elucidated non-contact thermal diodes using asymmetric nanostructures while \citep{st-gelais_demonstration_2014,shen2018near} reported on Near field RTR in the deep sub-wavelength regime between planar surfaces separated by nanometre-scale distances. Zhou et. al\citep{zhou2019radiation} proposed for instance a thermal diode using a nanoporous plate and a plate, both made of silicon carbide and where the two terminals are separated by a nanometric vacuum gap while Ott et. al\citep{ott2019radiative} explored  a radiative thermal diode made of two nanoparticles coupled with the nonreciprocal surface modes of a magneto-optical substrate.   

On the other hand, far-field (FF) radiative heat transfer also enables RTR and has  been an active area of interest in the past decade. Compared to NF RTR, FF RTR provides the advantage of simplicity and ease of fabrication since it does not require the combination and alignment of objects separated by micro-metric or nano-metric gaps.
Two main solutions have been reported for this purpose : the use of phase change materials (PCM) \citep{joulain2015modulation,nefzaoui_radiative_2014,ordonez2017photonic} and tunable meta-metrials often used as selective radiation emitters and absorbers\citep{nefzaoui_simple_2014}.  The use of phase change materials, and \(VO_2\) in particular, is predominant in both conceptual \citep{audhkhasi2019design} and experimental works \citep{ito2014experimental,jia2018far}. Recently, Ghanekar et. al\citep{ghanekar2017high} proposed the concept of a far-field radiative thermal rectifier combining both PCM, \(VO_2\) for instance, and a Fabry-Perot cavity based meta-material.
Beyond thermal rectification, various radiative heat transfer control devices have been proposed such as  thermal transistors \citep{prod2015dynamical,ordonez2019radiative,ben2014near,joulain2015modulation, joulain2016quantum} and photonic thermal memristors \citep{Ordonez_2019_Memristor} .  
The large majority of  those PCM based far-field radiation control devices use \(VO_2\) since the seminal work of Van Zwol et. al\citep{van2011phonon}  who provided a comprehensive characterization of \(VO_2\) radiative and thermo-radiative properties (TRP) around its metal-insulator transition temperature. Surprisingly, the good performances of \(VO_2\) for RTR seem to have prevented authors from exploring other materials, contrarily to the what have been done and reported for NF RTR. Indeed, the large variety of existing materials have not been systematically considered to assess their potential use in FF RTR. This can be legitimately explained by the scarcity of literature on materials’ radiative properties in the wavelength range of interest for thermal radiation and their subsequent temperature dependence. However, a thorough investigation of commonly available materials performances with respect to thermal rectification has not been realized up until now. 

In the present work, we present an evaluation of RTR potential of different materials commonly used in micro-fabricated devices such as Indium Arsenide, Gallium Arsenide, Gallium Antimonide, Germanium, Zinc Sulphide, Silicon and metals such as Copper and Gold based on radiative and thermo-radiative properties available in literature. The paper is organized as follows: in the present section, we introduce the concept of thermal rectification along with a brief review of the works accomplished so far and delineate the pertinence of the present work; next in section 2 , we explain the principle of radiative thermal rectification  and introduce the main quantities that govern and characterize a radiative thermal rectification behavior. We also, describe the used radiative and thermo-radiative properties of the considered materials and the process of relevant data extraction from literature. Finally, we report and discuss the main results on the radiative thermal rectification potential of the considered materials in section 3. 
\section{Methods}
\subsection{Theory}\label{theory}
\begin{figure*}[t!]
      \centering
      \includegraphics[scale=0.7]{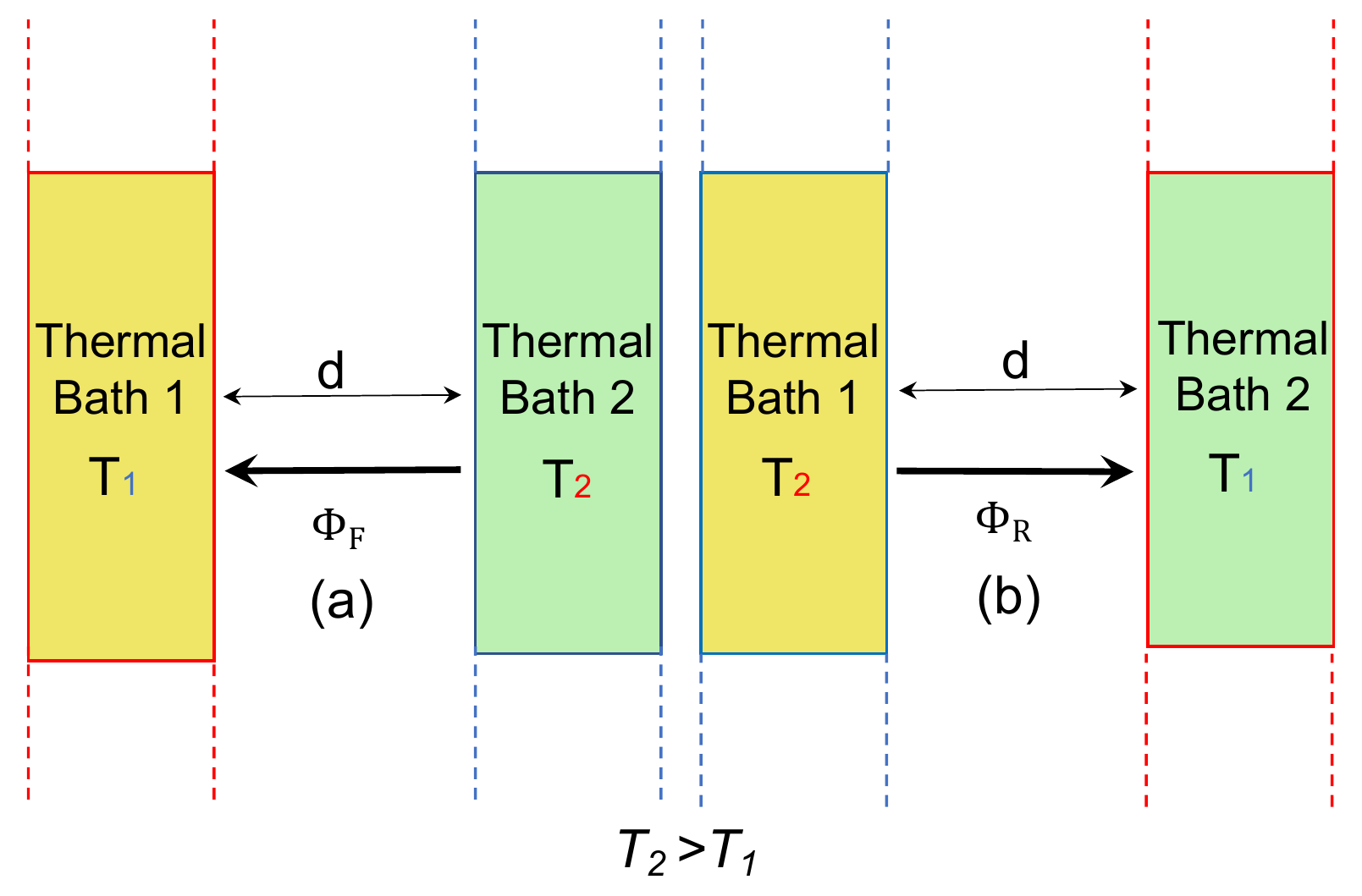}    
     \caption{\textmd{(a)Forward and (b) Reverse bias conditions of HT between two semi-infinite planes acting as thermal baths separated by a vacuum gap of thickness, d. Thermal rectification occurs only when, \(\Phi_{F}\) and \(\Phi_{R}\)  denoting the heat flux in the forward and reverse operating modes under two temperatures (T\(_{2}\) \textgreater T\(_{1}\)), are unequal. The gap width (d) is assumed to be much larger than the dominant thermal radiation wavelength (Wien wavelength, \(\lambda_{W}(T)\)).}}
      \label{fig:therm}
\end{figure*}

Variation of the heat flux magnitude when the sign of the temperature gradient between two points is reversed is the basic definition of thermal rectification. Let us consider two thermal reservoirs 1 and 2 at two different temperature \(T_ 1\)  and \(T_2\), respectively, with \(T_2 > T_1\). The temperature difference between the two reservoirs is noted \(\Delta{T}=T_2-T_1\). This temperature difference results in heat flux \(\Phi_{F}\). We will refer to this initial configuration as forward bias configuration. If we invert the temperature difference between the two reservoirs, i.e if we set reservoir 1 temperature to \(T_2\) and reservoir 2 temperature to \(T_1\), a reverse bias flux \(\Phi_R\) is observed. Thermal rectification occurs when the heat fluxes in forward \(\Phi_{F}\)  and reverse bias \(\Phi_{R}\), under the same magnitude of the thermal potential difference, i.e for a constant \(|\Delta{T}|\) but with opposite \(\Delta T\) signs, are unequal. Thermal rectification is generally quantified, even though it is not the only indicator encountered in the literature, by means of the normalized rectification coefficient \citep{starr_copper_1936, roberts_review_2011} which can be defined as: 
\begin{equation}\label{eq:RectificationDef}
     R =\frac{|\Phi_{F}-\Phi_{R}|}{max(\Phi_{F},\Phi_{R})}
\end{equation}
where, \(\Phi_{F}\) and \(\Phi_{R}\)  denote the heat flux under forward and backward bias, respectively. In this paper, the definition of normalized rectification coefficient given in equation \ref{eq:RectificationDef}  is preferred because it provides bound rectification coefficient values between zero and one, as opposed to alternative definitions adopted by some authors\citep{sawaki2011thermal, giazotto2013thermal} which lead to values from zero to infinity. 
A schematic of two thermal reservoirs exchanging heat flux in forward and reverse bias configurations is given in \ref{fig:therm}. 

In this paper, we consider the simple case of far-field radiative thermal rectification between two semi-infinite planes, separated by a  vacuum gap of thickness d, characterized by their radiative optical properties (their temperature dependent emissivities and reflectivities). Therefore, the distance separating the two planes exchanging heat by radiation is much larger than the characteristic wavelength of thermal radiation at the considered bodies temperatures given by Wien's law :  d\textgreater\(\lambda_{W}(T)\), where \(\lambda_{W}(T)\) = \(\frac{hc^{2}}{5k_{B}T}\), where h, c, k\(_{B}\) and T are Planck’s constant, the speed of light in vacuum, Boltzmann constant and the absolute temperature respectively. For simplicity’s sake, we also assume they are both Lambertian sources, meaning that their emissivities and reflectivities ( \(\epsilon(\lambda,T)\) and \(\rho(\lambda,T)\)  respectively) are direction independent\citep{modest2013radiative}.
Considering these assumptions, the net radiative heat flux (RHF) exchanged by the two bodies resumes then to the far field contribution, which can be written as\citep{joulain_radiative_2015}:
\begin{equation}\label{second_equation}
     \Phi(T_{1},T_{2})=\pi \int_{\lambda=0}^{\infty} [I^{0}(\lambda,T_{1})I^{0}(\lambda,T_{2})]\tau(\lambda,T_{1},T_{2})d\lambda
\end{equation}
where, 
\begin{equation}\label{third_equation}
     I^{0}(\lambda,T)=\frac{hc^{2}}{\lambda^5}\frac{1}{e^{\frac{hc}{k_{B}T}}-1}
\end{equation}
is the black body intensity at a temperature T and wavelength \(\lambda\), where h, c, k\(_{B}\) and T are Planck’s constant, the speed of light in vacuum, Boltzmann constant and the absolute temperature respectively\citep{bose_plancks_1924} and 
\begin{equation}\label{fourth_equation}
   \tau(\lambda,T_{1},T_{2})= \frac{\epsilon(\lambda,T_{1})\epsilon(\lambda,T_{2})}{1-\rho(\lambda,T_{1})\rho(\lambda,T_{2})}
\end{equation}
is the monochromatic RHF density transmission coefficient between 1 and 2, where \(\epsilon(\lambda,T)\)  and \(\rho(\lambda,T)\) is the monochromatic emissivity and reflectivity at a given temperature respectively. 
The RHF density transmission coefficient is governed by the radiative properties of the two considered thermal baths, in particular their emissivities and reflectivities  and the temperature dependence of these properties. These properties are completely governed by the considered bodies dielectric functions and geometries. 
In the case of opaque bodies, energy conservation\citep{modest2013radiative} and Kirchhoff’s laws \citep{modest2013radiative} combination leads to the following relation between the monochromatic emissivity, reflectivity and transmittivity at a given temperature:
\begin{equation}\label{fifth_equation}
  \epsilon(\lambda,T)=1-\rho(\lambda,T)-t(\lambda,T)
\end{equation}
Therefore, to evaluate the thermal rectification potential of a given pair of materials, we need to know the emissivity, reflectivity and transmittivity of each terminal of the considered thermal diode and its temperature dependence. Since the emissivity and reflectivity depend on the dielectric function, knowing the dielectric permittivity or the complex refractive index of the considered materials is also sufficient to estimate their potential for thermal rectification. In our case, these data have been harvested from literature.

\subsection{Radiative properties data extraction}
Thermo-radiative properties, i.e materials radiative properties and their temperature dependence, are the main input data required to assess the radiative thermal rectification ability of a given material. Difficulty to obtain such data, especially in the wavelength range of thermal radiation from room temperature to a few hundred Kelvins above room temperature, i.e. in the mid and far-infrared, perhaps explains the dearth of literature on thermal rectification with the large variety of existing materials and the strong concentration of literature on VO\(_{2}\). Although the most important material libraries in this regard such as The Handbook of Optical Materials by Edward D. Palik \citep{palik_handbook_1997} and that of Weber\citep{weber2002handbook}, provide an exhaustive collection of material optical properties, unfortunately, available data is not temperature dependent. Radiative properties of common materials at different temperatures, in the required infrared range, are scantily available in literature. Incidentally, the majority of relevant data found, regarding the works dedicated to this sub-area of research, are pertaining to the study of semiconductors. The optical properties of metals such as, Gold \citep{beran_reflectance_1985,aksyutov_temperature_1977}, Aluminum \citep{ujihara_reflectivity_1972,de_vries_temperature_1988}, Tungsten \citep{aksyutov_temperature_1977}, Molybdenum\citep{veal1974optical}, Silver \citep{ehrenreich1962optical}, Copper \citep{setien2014spectral}, Nickel \citep{setien2014spectral} have been studied, but unfortunately researchers have not been particularly interested in exploring the aforesaid optical properties at higher temperatures and at larger wavelengths, adherent to the IR range, which is an imperative in this study. Materials like Si \citep{francoeur_electric_2011} and Ge \citep{esaki1953properties} have been considered broadly, beginning in the 1950's. GaAs, InAs, InP, and GaSb make a case for volumes of research of their own and have been utilized as a part of light emanating and optoelectronic devices \citep{harris2010optical}. Material properties were learned at temperatures extending from low temperatures (liquid helium, 4.2 K, or liquid nitrogen, 77 K) to about room temperature, 298 K \citep{harris2010optical}. The 2010 work of Thomas R. Harris \citep{harris2010optical} provides valuable experimental data on the study of optical properties of Germanium, Gallium and Indium derivatives and on Bulk Silicon, at various temperatures. Optical properties of Silicon Carbide \citep{ng_temperature-dependent_1992,pitman_optical_2011}, and on Zinc Sulphide \citep{harris_thermal_2008} have also been reported in the past decades. The use of numerical simulation data for bulk silicon generated using a Drude Model was also suggested in \citep{chen2008heavily,basu_near-field_2013}, and we took use of this model to produce temperature dependent radiative properties data for bulk silicon. The list of materials considered in the present study is indicated in Table \ref{Mat}. 
\begin{table}[h!]
\begin{center}
\caption{Considered Materials}
\label{Mat}
\begin{tabular}{c|c|}
 \hline
 \multicolumn{1}{c}{ Materials} & \multicolumn{1}{c}{ Symbols}\\
 \hline
\multicolumn{1}{c}{Germanium\citep{harris2010optical,li1980refractive}} & \multicolumn{1}{c}{Ge}\\
\multicolumn{1}{c}{Gallium Derivatives\citep{harris2010optical,skauli2003improved}} & \multicolumn{1}{c}{GaAs,GaSb}\\
\multicolumn{1}{c}{Undoped Indium Arsenide\citep{harris2010optical,gillen2008temperature,bertolotti1990temperature}} & \multicolumn{1}{c}{InAs}\\
\multicolumn{1}{c}{Zinc Sulphide \citep{harris_thermal_2008,li1984refractive,leviton2013temperature,hawkins2004temperature}} & \multicolumn{1}{c}{ZnS}\\
\multicolumn{1}{c}{Silicon Carbide\citep{joulain_radiative_2015,ng_temperature-dependent_1992}} & \multicolumn{1}{c}{SiC}\\
\multicolumn{1}{c}{Bulk Doped N-type Silicon(\(1\times10^{20} cm^{-3}\))} & \multicolumn{1}{c}{DBuSi\(_{1}\)}\\
\multicolumn{1}{c}{Bulk Doped N-type Silicon(\(3\times10^{20} cm^{-3}\))} & \multicolumn{1}{c}{DBuSi\(_{3}\)}\\
\multicolumn{1}{c}{Gold \citep{beran_reflectance_1985,aksyutov_temperature_1977}} & \multicolumn{1}{c}{Au}\\
\multicolumn{1}{c}{Copper\citep{setien2014spectral}} & \multicolumn{1}{c}{Cu}\\
\hline
\end{tabular}
\end{center}
\end{table}

In the work of Thomas R. Harris\citep{harris2010optical}, temperature dependent transmission measurements for Ge were taken up to approximately 650 K. The data was taken in near IR and mid IR wavelength ranges. Comparison with older literature\citep{macfarlane_fine_1957} shows good agreement, indicating that the measured data are reliable. Similarly, temperature dependent transmission measurements were taken up to approximately 850 K for undoped GaAs and up to approximately 550 K for n-type GaAs doped with silicon. In the case for undoped GaSb, measurements were taken up to 650 K. Unfortunately, for sulphur doped InP, only near IR measurements were reported, which renders the data set incomplete for usage in the present study\citep{harris2010optical}. 
\begin{figure*}[t!]
      \centering
      \includegraphics[scale=0.65]{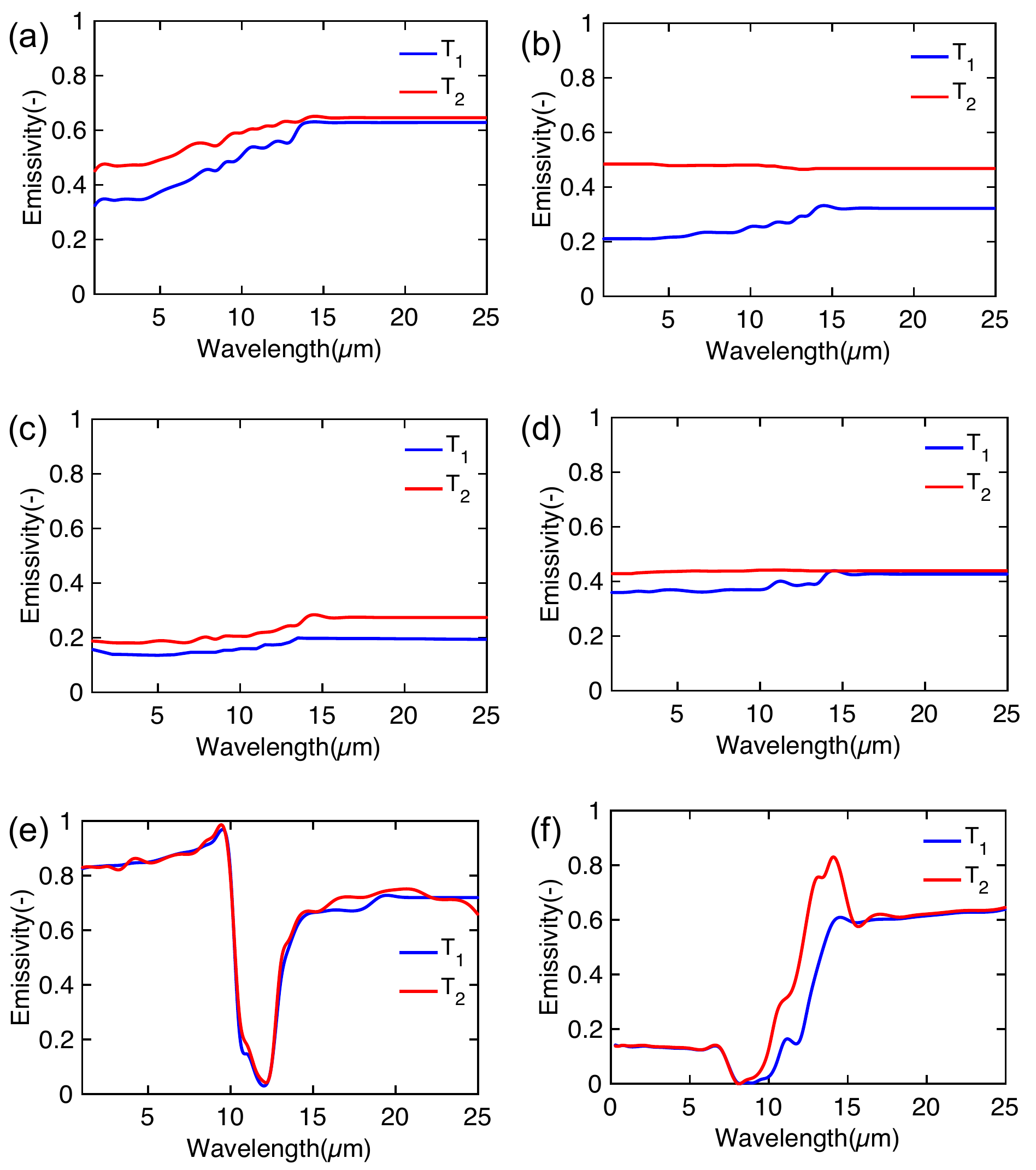}    
     \caption{\textmd{Emissivity of a few common materials found in literature-(a) GaSb (b) InAs (c) GaAs (d) Ge (e) SiC (f) ZnS at T\(_{1}\)=300K and T\(_{2}\)=500K.}}
      \label{fig:ref}
\end{figure*}
Thus, temperature dependent transmission measurements were carried out for Si, Ge, GaAs, GaSb, InAs, and InP from 0.6 to 25 \(\mu\)m at temperatures ranging from 295 up to 900 K. Band gap shifts were observed as temperature changed and were compared to previous works. General agreement was observed in the trend of the change in the band gap with temperature, however, the actual band gap energy values deviated from the expected-on average by about 10 \%. The reflectivity maximum increased in magnitude with increasing temperature, with successful measurements being done up to 517 K\citep{harris2010optical}.\newline

To completely gather temperature dependent reflectivity data of GaAs, GaSb, InAs, Ge another set of works were adhered to such as the work of Skauli et. al\citep{skauli2003improved}, where the refractive index of GaAs has been measured. From 1979-1984, H. H Li\citep{li1980refractive,li1984refractive} prepared a comprehensive report on the temperature dependence of the refractive indices of Si, Ge, ZnS, ZnSe, ZnTe etc., where he provided near non-existent temperature-dependent data on these particular semi-conductors refractive index in a wide spectral range including mid and far infra-red.

For SiC in the IR wavelength range from 1 \(\mu\)m to 25 \(\mu\)m, many data are available in literature and because much of SiC optical properties depend on its crystallographic orientation \citep{spitzer_infrared_1959,spitzer_infrared_1959-1}, they have been compared to other semiconductor relevant data with care. Some widely citepd sources are that of Bohren et. al \citep{bohren_absorption_2008}, Mutschke et al.\citep{mutschke_infrared_1999,clement_new_2003}, the measurements of Daniel Ng \citep{ng_temperature-dependent_1992} and Joulain et. al \citep{joulain_radiative_2015}. Cagran et. al\citep{cagran2007temperature} provided a comprehensive study on the temperature-resolved infrared spectral emissivity of SiC and this data was finally employed in the present study due to its reportage of wide and flexible temperature dependence of emissivity.

On the other hand, the measurements on Zinc Sulphide (ZnS) reported in \citep{harris_thermal_2008,leviton2013temperature,hawkins2004temperature} provide valuable insight into the thermal, structural and optical properties of a commercially available sample of multispectral ZnS. The hemispherical transmission, reflectivity, absorptivity, emissivity of unpolarized light at normal incidence on ZnS at 24 \degree C, 100 \degree C and 200 \degree C are reported in \citep{harris_thermal_2008}. 

\section{Results and discussion}
\begin{figure*}[t!]
      \centering
      \includegraphics[scale=0.60]{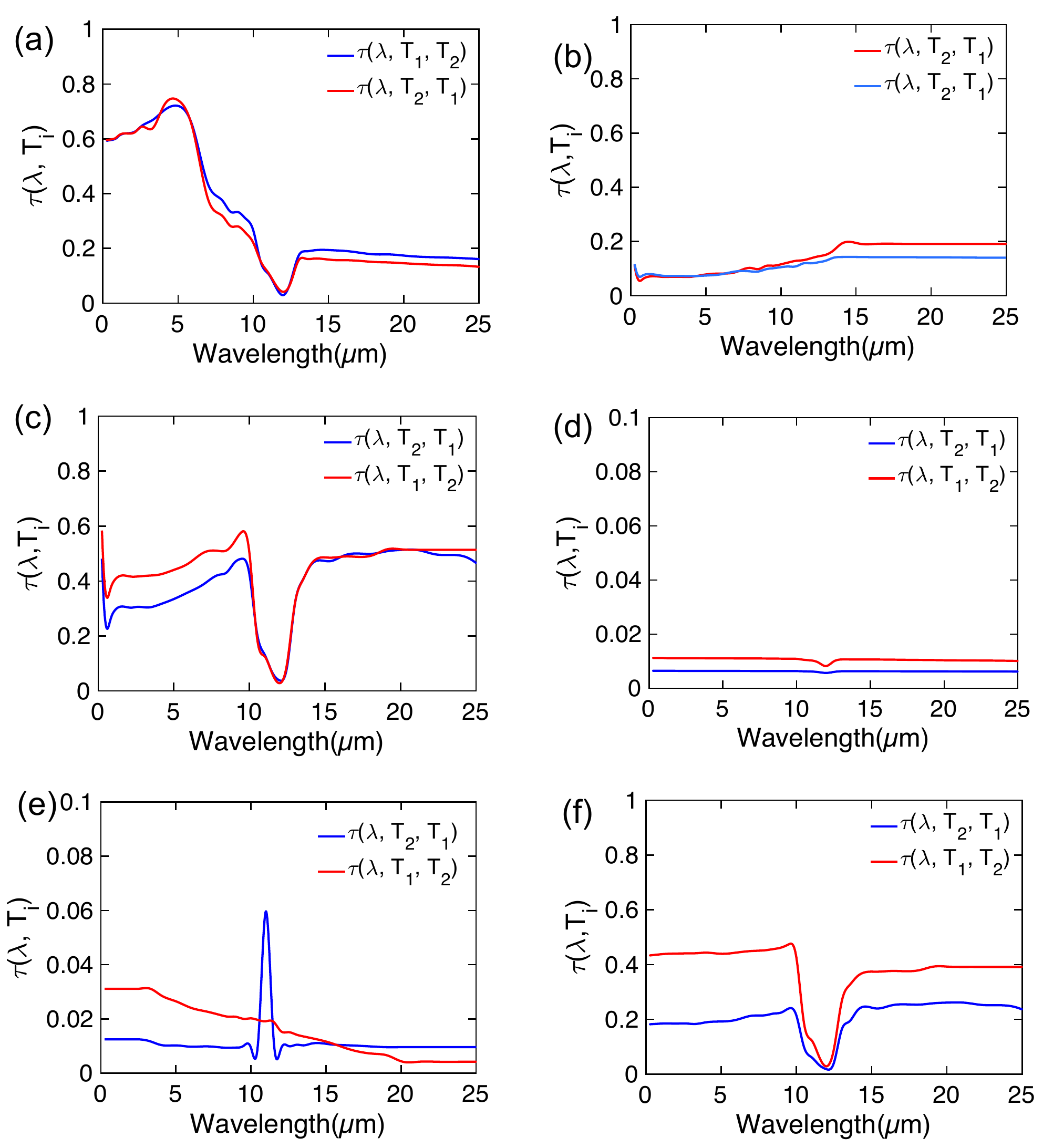}    
     \caption{\textmd{Obtained RHF density transmission coefficients in forward and reverse bias configurations between two materials- (a) SiC/DBuSi(\(1\times10^{20} cm^{-3}\)) ; (b) GaAs/GaSb; (c) SiC/GaSb; (d) SiC/Au; (e) InAs/Cu (f) InAs/SiC at T\(_{1}\)=300K and T\(_{2}\)=500K.}}
      \label{fig:tau}
\end{figure*}
A simple Matlab program based on equations \ref{eq:RectificationDef} to \ref{fifth_equation}, was implemented to quantify the thermal rectification coefficient for different pairs of materials, which takes the reflectivity data-set of two different materials and their temperature as input and calculates the forward and reverse bias fluxes, thereby finally giving the value of RTR coefficient as defined in equation \ref{eq:RectificationDef}. Radiative optical properties of the materials under consideration were studied at a temperature difference of 200 K, with  T\(_{1}\)=300K and T\(_{2}\)=500K while the spectral range taken into account is 1 - 25 \(\mu\)m. This temperature range was considered under the constraints of data availability. 

\begin{figure*}[t!]
      \centering
      \includegraphics[scale=0.50]{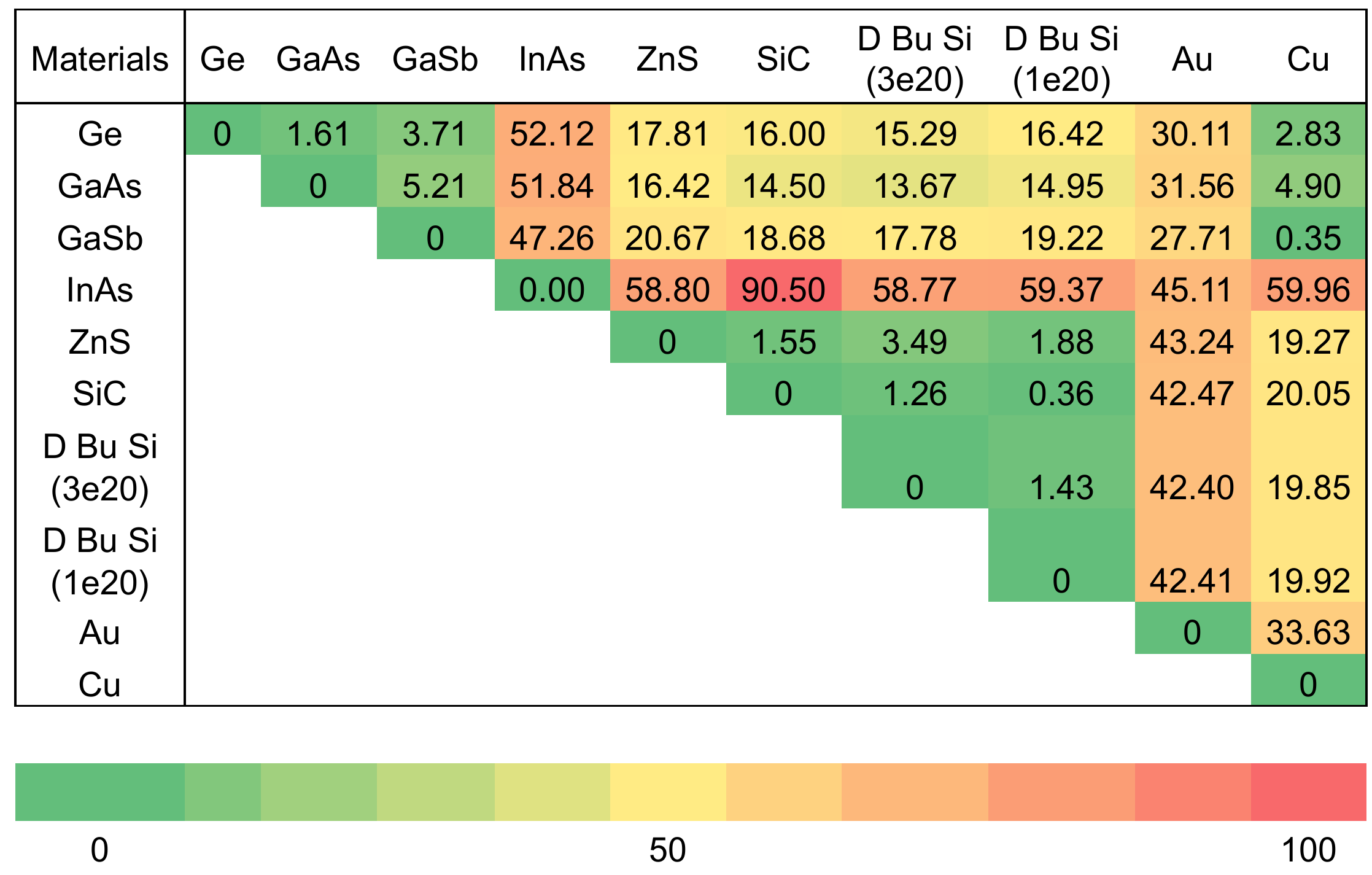}    
     \caption{\textmd{RTR coefficients of thermal rectifiers made of pairs of considered materials for a thermal bias of 200 K.}}
      \label{fig:rectmain}
\end{figure*}

Table\ref{Mat} shows the list of materials considered in this study. The respective unequal RHF density transmission coefficients in forward and reverse directions for considered pairs of materials are illustrated in Fig.\ref{fig:tau}, while Table \ref{fig:rectmain} summarises the obtained RTR coefficient for all possible pairs of materials with 44 pairs of materials considered in this study. 

Among the materials under consideration, Indium Arsenide (InAs) shows the largest values of RTR coefficient, up to 90.50\% when used with SiC, as shown in Table.\ref{fig:rectmain}. The highest value of RTR with materials other than InAs-SiC, can invariably be attributed to InAs-Cu which records a value of 59.96\%. As can be seen in Table.\ref{fig:rectmain}, all materials in combination with InAs record more than 50\% RTR except Au, which has an RTR value of 45.11\%. Au, in fact records greater 40\% with most materials in the table except Ge, GaAs and GaSb. Au-Ge, Au-GaAs combinations yield an RTR coefficient larger than 30\% and so does the metal-metal combination of Au-Cu. All other material combinations depicted in Table.\ref{fig:rectmain}, fall below 30\%.  The lowest value of RTR is recorded by a combination of GaSb-Cu and GaSb-DBuSi(1e20)- 0.5 \%.

\begin{figure*}[t!]
      \centering
      \includegraphics[scale=0.50]{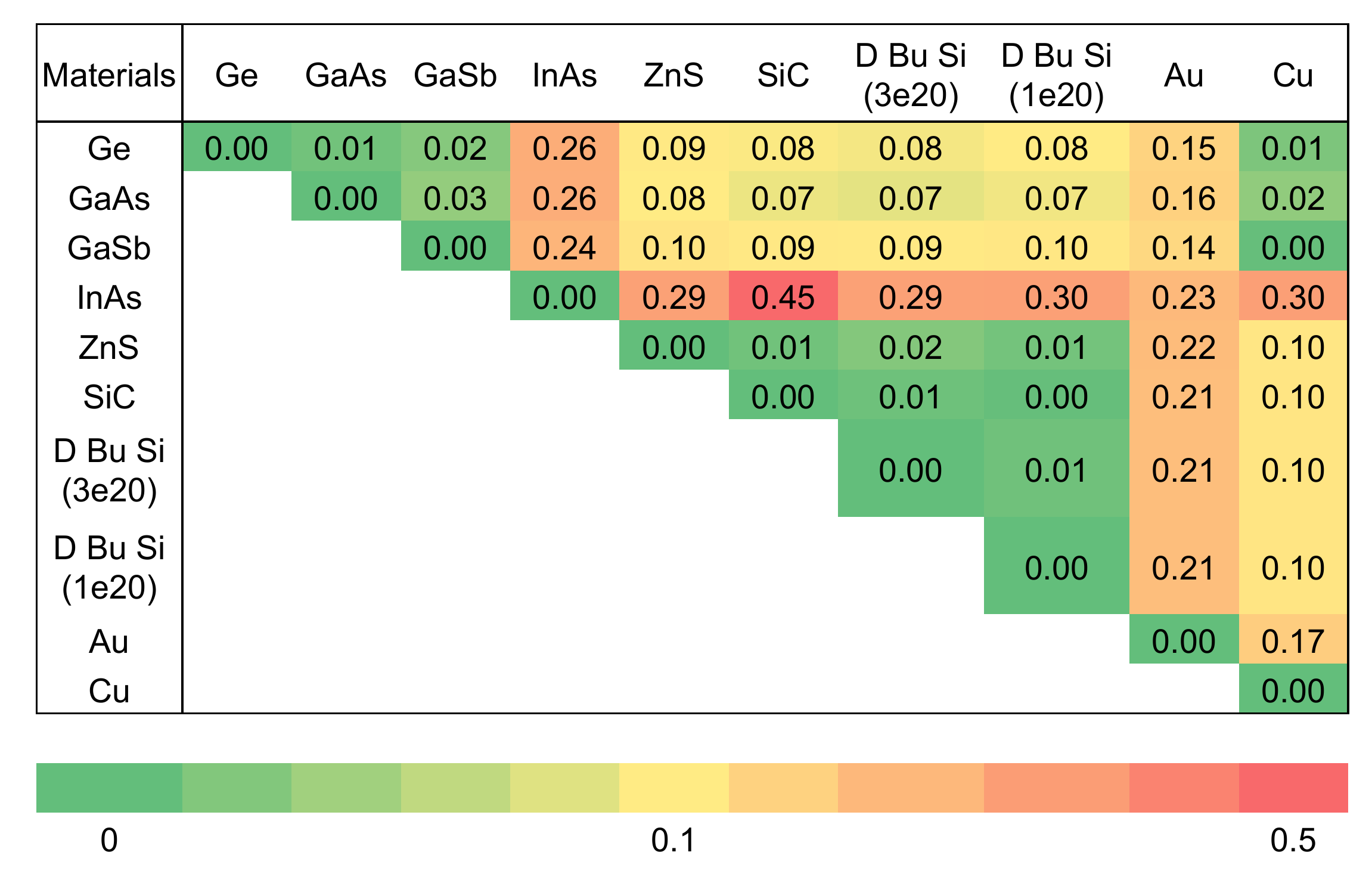}    
     \caption{\textmd{Normalized RTR coefficient i.e RTR coefficients of thermal rectifiers made of pairs of considered materials normalized by the thermal bias.}}
      \label{fig:rectK}
\end{figure*}
In order to obtain a perfect thermal rectifier, the spectra of the two thermal baths should perfectly match for a given temperature difference, so the exchanged heat flux is maximal, and perfectly mismatch when the thermal gradient is reversed so the exchanged flux is minimized. In fact, in the case of perfectly matching spectra, radiation emitted by one body at any wavelength is absorbed by the second body, while it is completely reflected and returned to the emitter in the case of perfectly mismatching spectra. A perfect one-way heat transmitter, i.e., a thermal diode can then be realised.

We show in Table \ref{fig:rectK}, the normalized RTR coefficient which is defined as a ratio of RTR to the magnitude of the considered thermal potential difference \(|\Delta{T}|\). Note that this definition of normalized RTR enables comparison of different pairs of materials operating under different thermal potential differences. However, this particular definition is not applicable for PCM materials which exhibit a significant change of their radiative properties with a very small temperature variation around their critical temperature which would lead to an infinite normalized RTR.

\begin{figure*}[hbt!]
      \centering
      \includegraphics[scale=0.4]{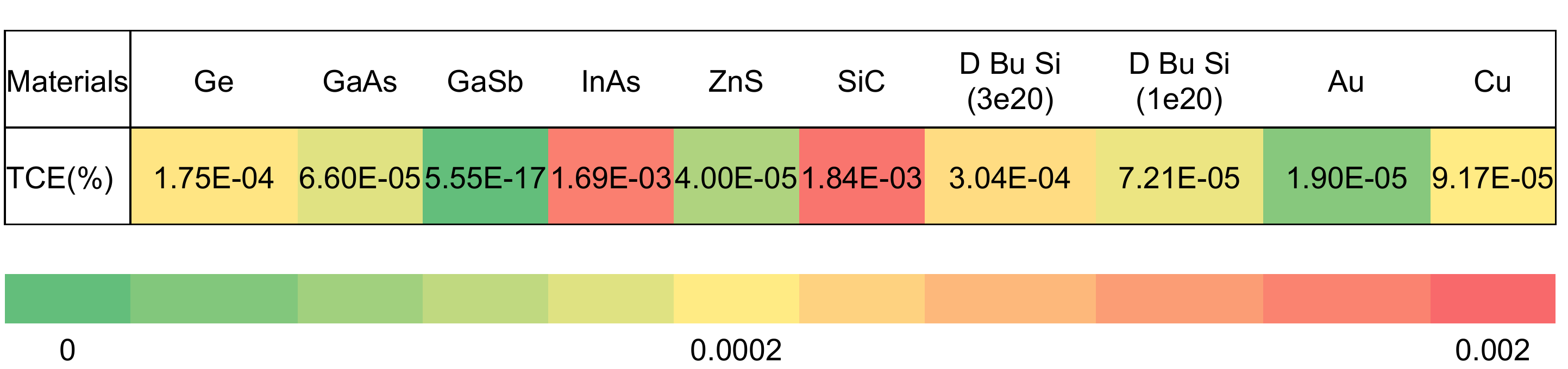}    
     \caption{\textmd{TCE(\%) of thermal rectifiers made of pairs of considered materials.}}
      \label{fig:tce}
\end{figure*}

In Table \ref{fig:tce}, we present the Temperature Coefficient of total Emissivity (TCE) that can be defined by \(\frac{d\epsilon_{t}}{dT}\), where, \(\epsilon_{t}\) is a spectral average with the spectral emissive power as a weighting factor and can be defined as \citep{modest2013radiative}: 
\begin{equation}\label{sixth_equation}
  \epsilon_{t}=\frac{1}{n^{2}{\sigma}T^{4}}\int_{0}^{\infty}\epsilon_{\lambda}(T,\lambda)E_{b\lambda}(T,\lambda)d\lambda
\end{equation}
where, \(\sigma\) is the Stefan–Boltzmann constant, T is the absolute temperature, \(\epsilon_{\lambda}(T,\lambda)\) is the spectral hemispherical emissivity and \(E_{b\lambda}(T,\lambda)\) is the blackbody spectral emissive power. This quantity(TCE) has been defined in a similar manner to the refractive index temperature coefficient, \(\frac{dn}{dT}\) commonly used in optics\citep{bertolotti1990temperature}. Here due to data availability constraints, a thermal bias of 200 K was considered. 
TCE can be used as a key indicator to identify the best candidate materials for thermal rectification since a large variation of radiative properties with respect to temperature is required to obtain a large thermal rectification coefficient. As can be observed in Table \ref{fig:tce}, the greater the TCE, the more susceptible the materials are for thermal rectification, thus reaffirming the best case of a far field radiative thermal rectifier composed of InAs and SiC. \newline
Results reported in the present manuscript could thus serve as red a basic guiding step for researchers and engineers interested in the design of radiative thermal rectifiers for the choice of best candidate materials. 
\section{Conclusion}
We reported in this paper on the thermal rectification potential of 44 pairs of  materials commonly used in several applications such as the microelectronics industry. At first, the theoretical concept of thermal rectification is introduced and the key quantities are defined. Then, used radiative properties, a key parameter for RTR coefficients calculation, are presented. Finally, results of the RTR coefficient of 44 pairs of materials are reported. Obtained results show that Indium Arsenide and Silicon Carbide can be very good candidates for a far-field radiative thermal rectifier when combined together with a rectification ratio up to 90.5\% for a thermal bias of 200 K.  We also show that several pairs of materials provide a relatively high rectification coefficient, larger than 50\% . This suggests that they can also be used for thermal rectifiers. A corollary of this result is that non negligible thermal rectification occurs when these pairs of materials are combined in a system with a thermal bias of the order of hundreds of kelvins. Taking into account the temperature dependence of the materials radiative properties is therefore mandatory for an accurate calculation of the exchanged heat flux between these materials in such systems. Presented results may be helpful for thermal management applications and in the advancement of the research and engineering of thermal rectifiers, thermal logical circuits and in thermal energy harvesting.
\section*{Acknowledgement}
This work was supported by SATT IDF Innov (now Erganeo) in the framework of the project DIOTHER and by the I-SITE FUTURE Initiative (reference ANR-16-IDEX-0003) in the framework of the project NANO-4-WATER.

\bibliography{jqsrt-sarkar-1}
\appendix

\newpage
\section*{Supplementary Materials I: Radiative properties extraction of considered materials}
\begin{table}[hbt!]
\begin{center}
\caption{Available data of considered materials}
\label{supp-tab:Mat}
\begin{tabular}{c|c|c|c|}
 \hline
 \multicolumn{1}{c}{Materials}&\multicolumn{1}{c}{Reflectivity(\(\rho\))} &\multicolumn{1}{c}{Transmittivity(t)}  &\multicolumn{1}{c}{Emissivity(\(\epsilon\))}\\
 \hline
\multicolumn{1}{c}{Ge} & \multicolumn{1}{c}{\citep{li1980refractive}} & \multicolumn{1}{c}{\citep{harris2010optical}} & \multicolumn{1}{c}{\(\epsilon\)=1-\(\rho\)-t}\\
\multicolumn{1}{c}{GaAs} & \multicolumn{1}{c}{\citep{gillen2008temperature,skauli2003improved}} & \multicolumn{1}{c}{\citep{harris2010optical}} & \multicolumn{1}{c}{\(\epsilon\)=1-\(\rho\)-t}\\
\multicolumn{1}{c}{GaSb} & \multicolumn{1}{c}{\citep{herve1995empirical}} & \multicolumn{1}{c}{\citep{harris2010optical}} & \multicolumn{1}{c}{\(\epsilon\)=1-\(\rho\)-t}\\
\multicolumn{1}{c}{InAs} & \multicolumn{1}{c}{\citep{bertolotti1990temperature,gillen2008temperature}} & \multicolumn{1}{c}{\citep{harris2010optical}} & \multicolumn{1}{c}{\(\epsilon\)=1-\(\rho\)-t}\\
\multicolumn{1}{c}{ZnS} & \multicolumn{1}{c}{\citep{harris_thermal_2008,li1984refractive,leviton2013temperature,hawkins2004temperature}} & \multicolumn{1}{c}{\citep{harris_thermal_2008}} & \multicolumn{1}{c}{\citep{harris_thermal_2008}}\\
\multicolumn{1}{c}{SiC} & \multicolumn{1}{c}{\(\rho\)=1-\(\epsilon\)} & \multicolumn{1}{c}{opaque sample\citep{cagran2007temperature}} & \multicolumn{1}{c}{\citep{cagran2007temperature}}\\
\multicolumn{1}{c}{Au} & \multicolumn{1}{c}{\citep{aksyutov_temperature_1977}} & \multicolumn{1}{c}{opaque sample\citep{aksyutov_temperature_1977}} & \multicolumn{1}{c}{\(\epsilon\)=1-\(\rho\)}\\
\multicolumn{1}{c}{Cu} & \multicolumn{1}{c}{\citep{setien2014spectral}} & \multicolumn{1}{c}{opaque sample\citep{setien2014spectral}} & \multicolumn{1}{c}{\citep{setien2014spectral}}\\
\hline
\end{tabular}
\end{center}
\end{table}
Radiative optical properties sourced from the works citepd in this paper, have provided all the necessary temperature dependent data as is required in this study. Table\ref{supp-tab:Mat} delineates from which reference the sample radiative optical properties have been gathered and how a particular material property, if not available, of each sample has been calculated. Here, recalling the notations in section \ref{theory}, \(\epsilon(\lambda,T)\), \(\rho(\lambda,T)\) and \(t(\lambda,T)\) are the material emissivity, reflectivity and transmittivity  respectively.  

\newpage
\section*{Supplementary Materials II:  Mapping RTR coefficient of considered materials}
\begin{figure*}[t!]
      \centering
      \includegraphics[scale=0.23]{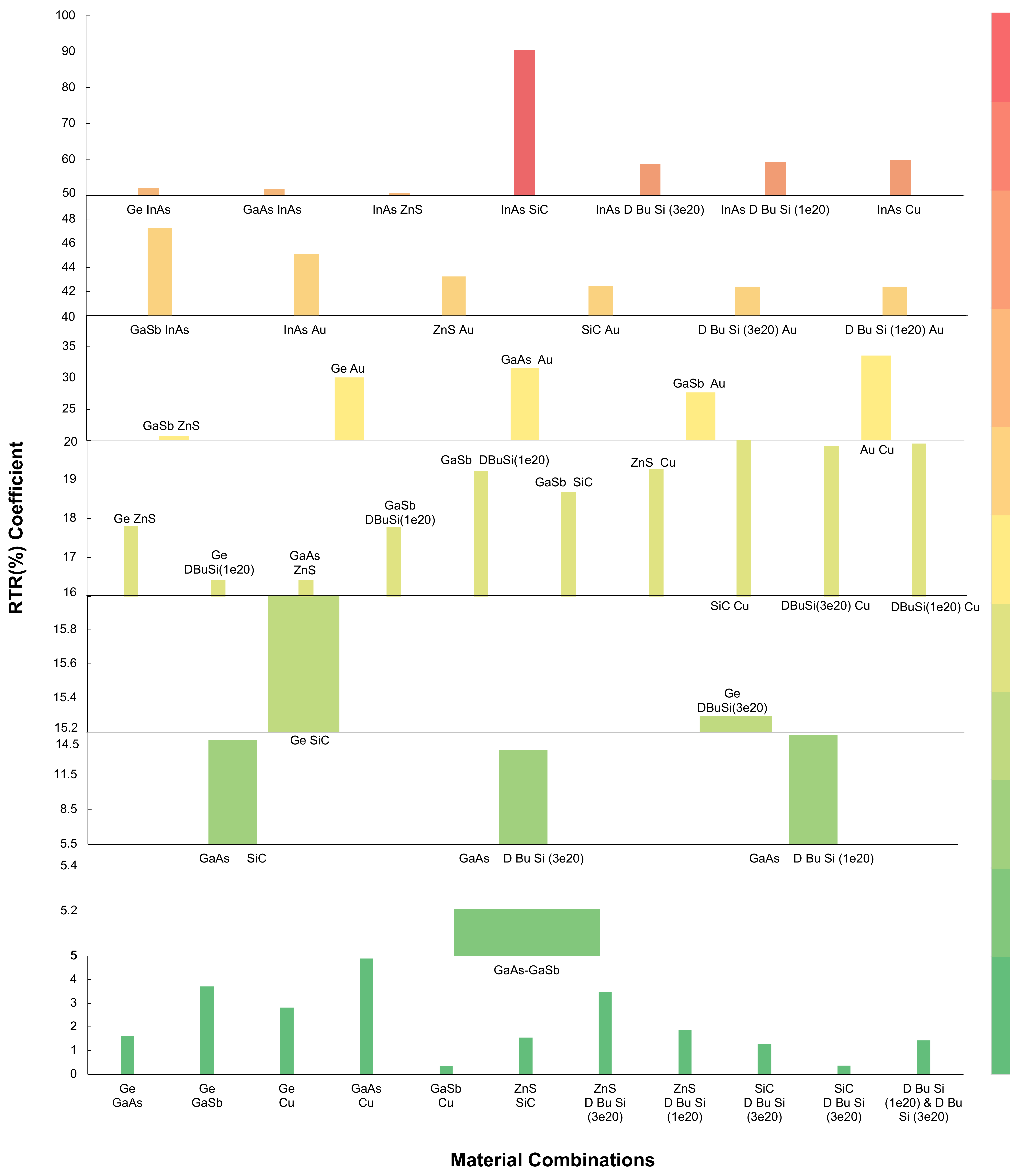}    
     \caption{\textmd{Mapping RTR coefficients of thermal rectifiers made of pairs of considered materials for a thermal bias of 200 K based on the table in Fig.\ref{fig:rectmain}}}
      \label{fig:rect}
\end{figure*}
As depicted in Table \ref{fig:rectmain}, RTR coefficients for different combinations of considered materials, belong to different ranges from 0-91\%. In an attempt to streamline and to make each material combination identifiable for an appropriate RTR coefficient, all RTR coefficients are mapped in Figure\ref{fig:rect} in different ranges namely: 0-5\%,  5-5.5\%,  5.5-15.2\%, 15.2-16\%, 16-20\%, 20-40\%, 40-50\%, 50-100\%. These ranges have been ascertained relying on the overall data range in Table \ref{fig:rectmain}. Each bar height represents the value of the RTR coefficient and the bar width represents the distribution of entries for each data range. The bar color map corresponds to the amplitude of RTR i.e the lowest values starting from green to red being the highest  value obtained. As seen in Figure\ref{fig:rect}, there are highest no of entries in the range from 16-20\% and 0-5\%.  There are 7 material combinations in the range of 50-100\%, which suggests that these combinations have non negligible thermal rectification. 

\end{document}